\documentclass[aps, pra, preprint, groupedaddress, amsfonts,
               amsmath, amssymb, nofootinbib]{revtex4-1}
\usepackage{microtype}
\usepackage{graphicx}
\usepackage{epstopdf}
\usepackage[utf8]{inputenc}
\usepackage[T1]{fontenc}
\usepackage[usenames,dvipsnames]{xcolor}
\usepackage{hyperref}

\usepackage{color}

\begin{document}
\title{Indication of strong interatomic Coulombic decay 
in slow He$^{2+}$-Ne$_2$ collisions}

\author{Tom Kirchner}  
\email[]{tomk@yorku.ca}
\affiliation{Department of Physics and Astronomy, York University, Toronto, Ontario M3J 1P3, Canada}
\date{\today}
\begin{abstract}
Electron removal in collisions of
alpha particles with neon dimers is studied using an independent-atom-independent-electron
model based on the semiclassical approximation of heavy-particle collision physics.
The dimer is assumed to be frozen at its 
equilibrium bond length and collision events for the two ion-atom subsystems are combined
in an impact parameter by impact parameter fashion for three 
mutually perpendicular orientations.
Both frozen atomic target and dynamic response model calculations are carried out
using the
coupled-channel two-center basis generator method.
We pay particular attention to inner-valence Ne($2s$) electron removal, which
is associated with interatomic Coulombic decay (ICD), resulting
in low-energy electron emission and dimer fragmentation.
Our calculations confirm a previous experimental result at 150 keV/amu
impact energy regarding the relative strength of ICD compared to direct
electron emission. They further
indicate that ICD is the dominant Ne$^+$ + Ne$^+$ fragmentation 
process	below 10 keV/amu,
suggesting that a strong low-energy electron 
yield will be observed in the ion-dimer system in a regime in which
the creation of continuum electrons is a rare event in the ion-atom problem.
\end{abstract}

\maketitle
\section{Introduction}
\label{intro}
The dominant reaction channel in low-energy ion-atom collisions is electron
capture by the charged projectile. 
For example, in the He$^{2+}$-Ne system the cross section for net capture plateaus 
at a value on the order of $10^{-15}$ cm$^2$ at impact energies 
$E_P \approx 10$ keV/amu,
while that for direct transfer to the continuum is about an order of magnitude smaller
and decreases rapidly toward lower $E_P$~\cite{tom00}, i.e. it is exceedingly
unlikely to find free electrons in this regime.

If the target atom happens to be in close proximity to another atom this may
change, since a free-electron-producing channel that is non-existent
in isolated atomic targets can open up:
interatomic Coulombic decay (ICD).
ICD involves the transfer of excitation energy, typically associated with
the formation of a hole in an inner-valence orbital, from one atom to a
neighboring one, and the release of that energy in the form 
of a low-energy continuum electron.
The latter feature is of considerable applied relevance, since low-energy
electrons are known to play an important role in numerous phenomena,
most notably perhaps in inflicting damage in biological tissue~\cite{Bouda2000}.

ICD has been first observed 
in photoexcited noble gas clusters~\cite{Marburger03, Jahnke04},
and has since been shown to occur in a variety of systems, including 
ion-impact collision problems at intermediate~\cite{Titze11, Kim11, Kim13}
and low $E_P$~\cite{Iskandar15}.
A number of review articles on ICD and its various forms have been published, 
among them~\cite{Jahnke15}
which focused on experimental studies and also discussed the relevance of
ICD in biological systems, and, more recently, the comprehensive paper~\cite{Jahnke20}.

The observation of ICD in low-energy ion-impact collisions was perhaps somewhat 
surprising, given that the dominant
capture processes usually involve outer-shell target electrons
whose removal does not generate excitation energy in the  
ion left behind.
In fact, in the experiments reported in~\cite{Iskandar15} 
evidence for the occurrence of ICD was found in only one out
of three collision systems studied, namely in 2.81 keV/amu
O$^{3+}$-Ne$_2$ collisions, while the kinetic energy release spectra obtained for
3.37 keV/amu Ar$^{9+}$ and 2.28 keV/amu Xe$^{20+}$ impact
on Ne$_2$
suggested that ICD plays no appreciable role in these cases.
Classical over-the-barrier model 
calculations published along with the data 
explained the absence of ICD for the Ar$^{9+}$ and Xe$^{20+}$ 
projectiles with strong Ne($2p$) capture happening
simultaneously with and effectively overwhelming the inner-valence electron removal 
processes required to trigger ICD.

In a recent work using a similar model as in the present paper,
we were able to confirm the occurrence of ICD for O$^{3+}$ projectiles
in a semi-quantitative fashion~\cite{Dyuman20}. We further showed that the situation
would be different and ICD very weak if bare triply-charged ions, i.e. 
Li$^{3+}$ had been used in the experiment. This is so because the energy level
structure of the bare projectile does not allow for efficient  
capture of a Ne($2s$) electron. 
Together with the findings of~\cite{Iskandar15} this suggested
that a delicate balance of
charge state and structure of a projectile is required to make
ICD a significant process. The question was raised if a projectile
can be identified that maximizes the ICD yield in low-energy capture collisions~\cite{Dyuman20}.

The main objective of the present work is to report on a promising candidate:
alpha particles, i.e. He$^{2+}$ ions. Bare (and non-bare) helium ions were
used in the experiments reported in~\cite{Kim11, Kim13}, but at 
higher impact energies $125$ keV/amu $ \le  E_P \le 162.5$ keV/amu, where direct target ionization 
competes with electron capture. ICD was successfully identified,
but for most situations studied the direct electron emission yield  
was larger than the yield associated with ICD. We confirm this finding for
He$^{2+}$ collisions at $E_P = 150$ keV/amu and show that the
situation is very different at $E_P \lessapprox 10$ keV/amu.

Our model is briefly summarized in Sec.~\ref{sec:model}. For a more detailed
discussion the reader is referred to our previous paper~\cite{Dyuman20}.
In Sec.~\ref{sec:results} we present and discuss our results and in
Sec.~\ref{sec:conclusions} we offer a few concluding remarks.

\section{Model}
\label{sec:model}

We work in the framework of the semiclassical approximation and assume that the
projectile ion travels on a 
straight-line trajectory with constant speed $v=|\bf v|$. The 
neon dimer is assumed to remain fixed in space during the collision
at its  equilibrium bond length 
$R_e=5.86$ a.u.~\cite{Cybulski99, Iskandar15},
and its interaction with the projectile is described as two independent
ion-atom collisions occurring at impact parameters $b^{A}$ and $b^{B}$  which, depending on
the relative orientation of the dimer with respect to the projectile
velocity vector $\bf v$, may be the same or different.
Three perpendicular orientations are
considered: One in which the projectile travels parallel to the dimer axis
such that $b^{A}=b^{B}=b$, where 
$b$ is the impact parameter measured with respect to
the center of mass of the dimer (orientation I), one in which the dimer 
is perpendicular to  $\bf v$ in the scattering plane defined by $\bf v$ and $\bf b$
(orientation II), and one in which the dimer is perpendicular to the
scattering plane.
Transition probabilities for the ion-dimer system are calculated
by combining probabilities for those ion-atom processes
at $b^{A}$ and $b^{B}$ which correspond to a given process of interest $j$ at the
impact parameter $b$, and an orientation average is constructed
for each $j$ according to  
\begin{equation}
	P_j^{\rm ave} (b) = \frac{1}{3}\left[ P_j^{\rm I}(b) +  P_j^{\rm II}(b) + P_j^{\rm III}(b) \right] .
	\label{eq:pave}
\end{equation}
Orientation-averaged total cross sections are obtained by integration over the impact parameter 
\begin{equation}
	\sigma_j^{\rm ave} = \int P_j^{\rm ave}(b) d^2 b = 2\pi \int_0^{b_{\rm max}} bP_j^{\rm ave}(b) db 
	\label{eq:tcs}
\end{equation}
with $b_{\rm max}=10$ a.u., which turned out to be sufficient to capture all appreciable
contributions. We note that
the same orientation-averaging procedure was used in numerous theoretical studies of
collisions involving hydrogen molecules (see, e.g.~\cite{Luehr10, Gao20} and references
therein). 

The ion-atom collision calculations are carried out at the level of the independent electron
model (IEM). We consider both a frozen target potential (no-response) model and a
more sophisticated model which includes dynamic response following the prescription
of~\cite{tom00}. 
The atomic neon potential used in both variants is obtained from an optimized potential method (OPM) 
calculation at the level of the exchange-only approximation~\cite{ee93}.
The gist of the response model is to adjust the potential during the collision
in a way that reflects the ionic character a target atom acquires when electrons are removed from it.
The time-dependent net electron removal probability is used to regulate the ionic character
such that response effects remain small as long as zero-fold and one-fold
electron removal dominate, and become appreciable once higher-order removal processes 
contribute substantially~\cite{tom00}. 
The response model has been applied successfully to numerous
ion-atom collision problems~(see, e.g.~\cite{tom04, Schenk15, leung17} and references therein), 
but one should bear in mind that it deals in a global fashion
with dynamic effects which ideally would be handled at a \textit{microscopic} level,
e.g. in a time-dependent OPM framework.
The latter is challenging to implement for atoms with more than two 
electrons~\cite{Keim03, Baxter17} and is beyond the scope of the present work.

Following our previous work for triply-charged ion-dimer collisions~\cite{Dyuman20}
we assume the Ne $K$-shell electrons
to be passive and solve the single-particle Schr\"odinger
equations for the eight Ne $L$-shell electrons using
the two-center basis generator method (TC-BGM).
The basis sets used include
the $2s$ to $4f$ target orbitals and 
(i) at impact energies $E_P \ge 20$ keV/amu all projectile orbitals
from $1s$ to $4f$, and (ii) all projectile orbitals
up to $7i$ at lower $E_P$. 
We have checked at $E_P = 10$ keV/amu that including these additional
projectile states result in insignificant changes except at impact
parameters smaller than one atomic unit.

The bound-state two-center basis sets are augmented by
pseudo states to account for couplings to the continuum
using the standard BGM procedure of operating with powers of a regularized projectile
potential operator on the target eigenstates~\cite{tcbgm}. 
For the low projectile energy calculations we included the same set of states as
in~\cite{Dyuman20}, while at $E_P \ge 20$ keV/amu a significantly larger
set of pseudo states, similar to the sets discussed in~\cite{Schenk15},
was used in order to describe
the increasing importance of transitions to the continuum with sufficient
accuracy. The latter reference includes basis set convergence
studies and also provides some details
on the implementation of the dynamic response model of~\cite{tom00}
within the TC-BGM.

The impact parameter by impact parameter combination of atomic results
to assemble transition probabilities for the ion-dimer system is based
on straightforward multinomial statistics involving products of single-electron
transition probabilities, i.e. it is consistent with the
IEM. We are not interested in a complete analysis including all
conceivable removal processes of a total of $N$ electrons, but focus on those which can
be associated with the Ne$^+$ + Ne$^+$ fragmentation channel.
These are 
(i) the removal of one $2p$ electron from each of the two neon atoms
resulting in the direct production of the Ne$^+(2p^{-1})$ + Ne$^+(2p^{-1})$
configuration, 
(ii) the similar direct production of the Ne$^+(2s^{-1})$ + Ne$^+(2p^{-1})$
configuration, 
(iii) the direct production of Ne$^+(2s^{-1})$ + Ne$^+(2s^{-1})$,
(iv) the removal of two $2p$ electrons from one of the two atoms while
the other atom remains neutral,
(v) the removal of one $2s$ electron from one of the two atoms while
the other atom remains neutral.
We will refer to processes (i) to (iii) as direct Coulomb explosion (CE)
channels\footnote{%
Processes (ii) and (iii) generate some excitation energy in one or both atoms, but
not enough to ionize an additional electron of the dimer, i.e. they cannot trigger
ICD.}.
Process (iv) leads to a transient state which can relax to the 
Ne$^+(2p^{-1})$ + Ne$^+(2p^{-1})$ configuration via radiative charge
transfer (RCT) and process (v) corresponds to ICD~\cite{Iskandar15, Dyuman20}. 

The explicit multinomial expressions for these processes are lengthy.
For the first CE process and the processes associated
with RCT and ICD they were given explicitly in~\cite{Dyuman20} for orientation I.
In this work we go one step further by separating all five processes in terms of transitions
to bound projectile states (capture) and the continuum (ionization). 
As an example, let us consider the $2s$-vacancy production process (v) associated with ICD.
In orientation I in which $b_A=b_B=b$ the probability for this process is
\begin{equation}
	P_{2s^{-1}}^{x}(b) =
	4 p_{2s}^{x}(b) [1-p_{2s}^{\rm rem}(b)]^3 [1-p_{2p_0}^{\rm rem}(b)]^4 
	[1-p_{2p_{1}}^{\rm rem}(b)]^8 ,
	\label{eq:picd}
\end{equation}
where $p_{2s}^x$ is the single-particle probability for $x=$ capture or $x=$ ionization from the
$2s$ initial state, and $p_i^{\rm rem}$ denote the single-particle removal
probabilities from the $i=\{2s, 2p_0, 2p_1 \}$ states. The factor of four reflects the fact that each of the four $2s$ electrons
of the dimer can be the one that is captured or ionized and the $(1-p_i^{\rm rem})$ terms
ensure that none of the other electrons is removed in the same event.
Note that due to the symmetry of the ion-atom problem with respect to reflections 
about the scattering plane the probabilities for the $2p_{+1}$ and $2p_{-1}$ initial states are
the same.
The separation into capture and ionization events allows
us to estimate the ratio of electron emission due to ICD compared to direct
emission which was measured in~\cite{Kim13} at $E_P = 150$ keV/amu.

\section{Results and discussion}
\label{sec:results}
\begin{figure}
\begin{center}$
\begin{array}{cc}
\resizebox{0.52\textwidth}{!}{\includegraphics{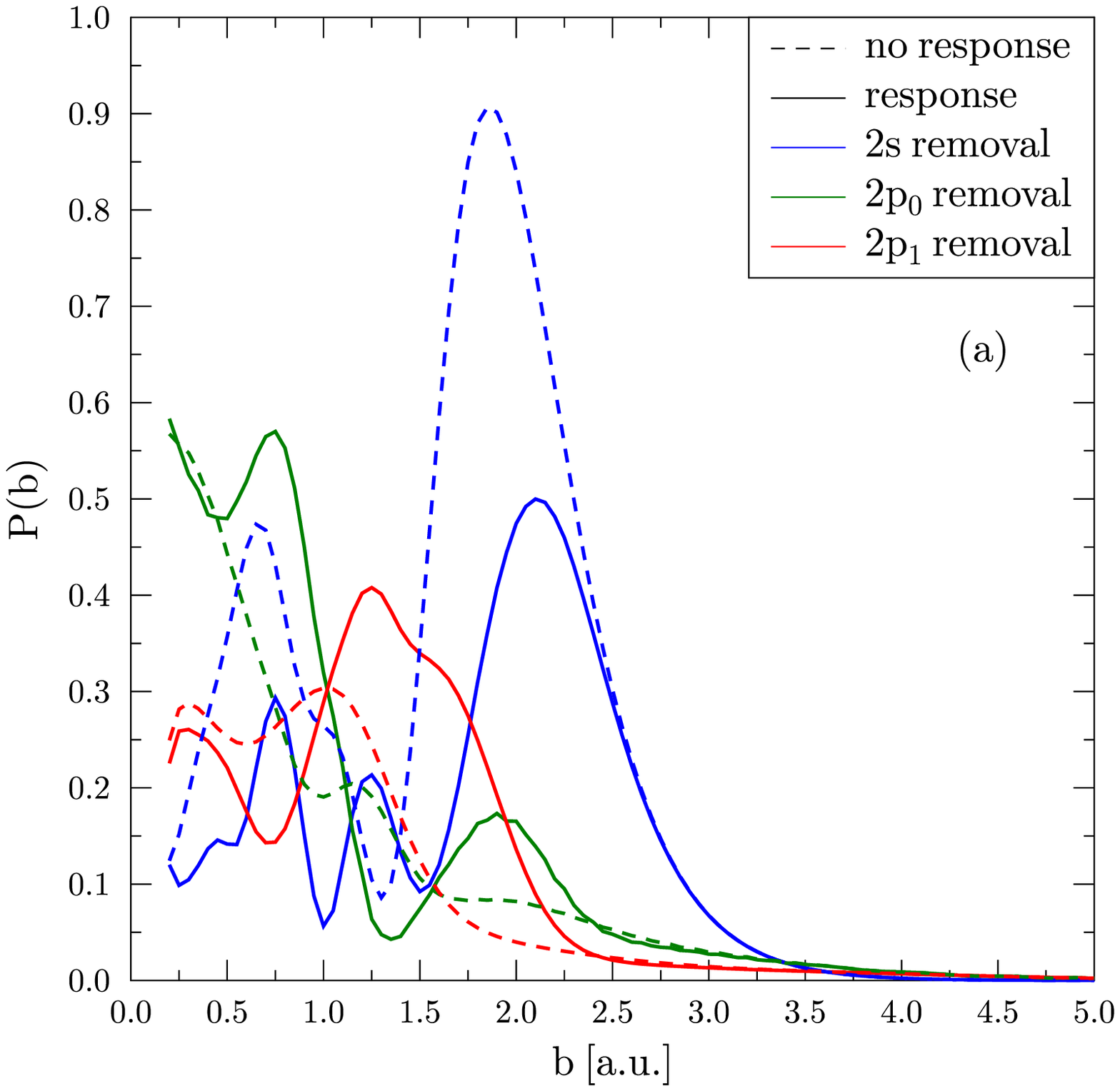}}&
\resizebox{0.52\textwidth}{!}{\includegraphics{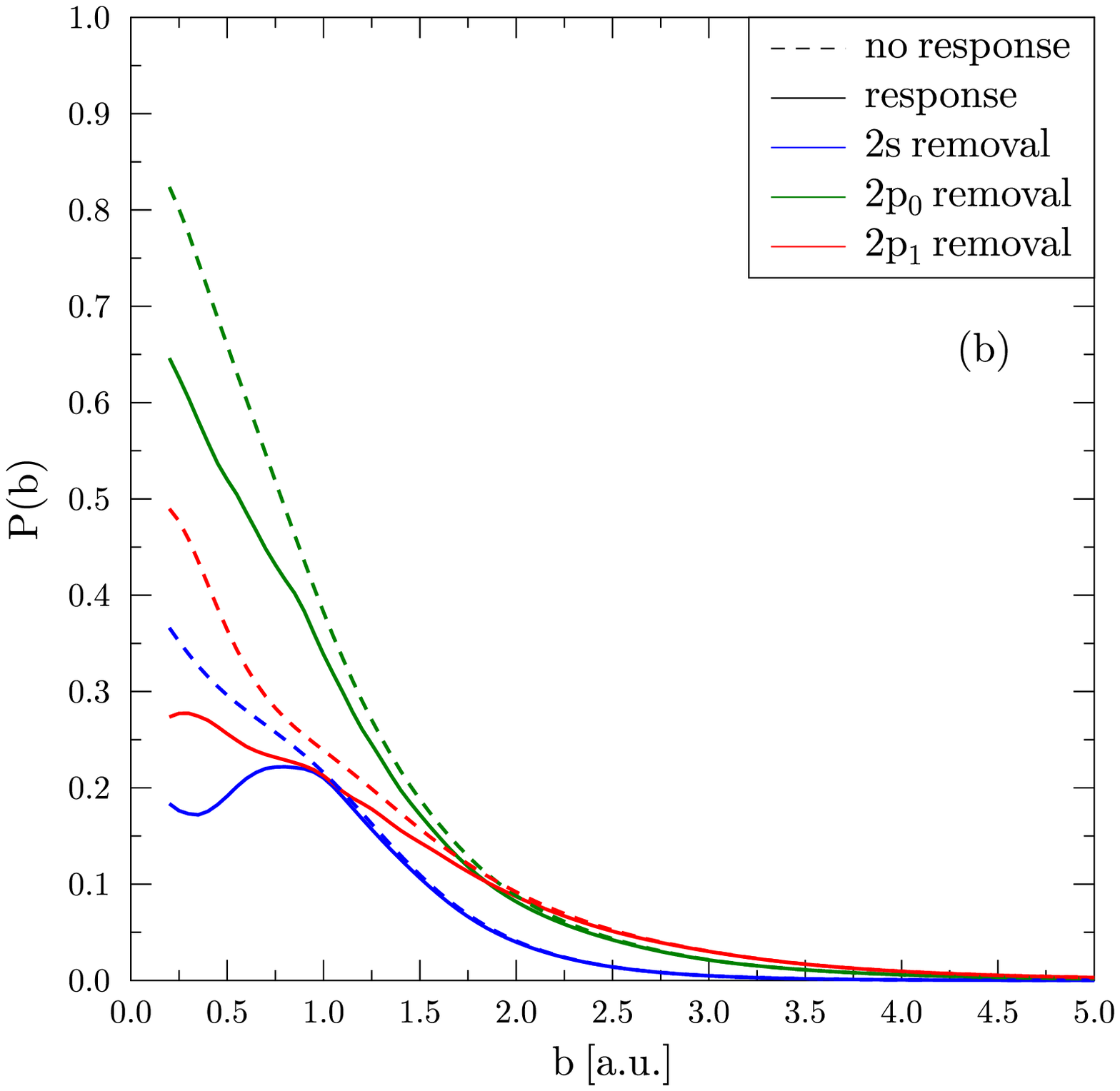}}
\end{array}$
\caption{%
Single-particle probabilities for electron removal from the Ne $L$ shell by 
(a) $E_P=10$ keV/amu and (b) $E_P=150$ keV/amu
He$^{2+}$ impact plotted as functions of the impact parameter.}
\label{fig1}
\end{center}
\end{figure}

We begin the discussion of results with a look at the atomic single-particle removal
probabilities obtained in both no-response and response models. Figure~\ref{fig1} shows these
probabilities at $E_P = 10$ keV/amu in panel (a) and at $E_P = 150$ keV/amu in panel (b).
The current results are consistent
with our earlier He$^{2+}$-Ne calculations reported in~\cite{tom98, tom00, Schenk15} (not shown).

Response effects are significant, particularly at the lower energy
where electron capture to the projectile ground state is the dominant removal process.
Only at relatively large impact parameters where all single-particle removal
probabilities tend to
be small do the no-response and response results merge. 
At $E_P = 10$ keV/amu removal from the
Ne($2s$) initial state is very strong, particularly
in the no-response model. This is due to the closeness of the 
single-particle energy eigenvalues of the initial 
($\varepsilon_{{\rm Ne}(2s)}^{\rm OPM}=-1.718$ a.u.)
and final ($\varepsilon_{{\rm He}^{+}(1s)}=-2$ a.u.) 
states and their associated coupling. Dynamic response appears to weaken
this coupling somewhat.
The initial Ne($2p$) electrons
are less likely to be captured because they are energetically farther
away ($\varepsilon_{{\rm Ne}(2p)}^{\rm OPM}=-0.851$ a.u.).

At $E_P = 150$ keV/amu transitions to the continuum dominate
and the no response vs. response comparison mostly
reflects the fact that ionization becomes less likely with
increasing binding energy. The latter is also the
reason why the $2p$ removal probabilities are larger than the $2s$
probability.

\begin{figure}
\begin{center}$
\begin{array}{cc}
\resizebox{0.52\textwidth}{!}{\includegraphics{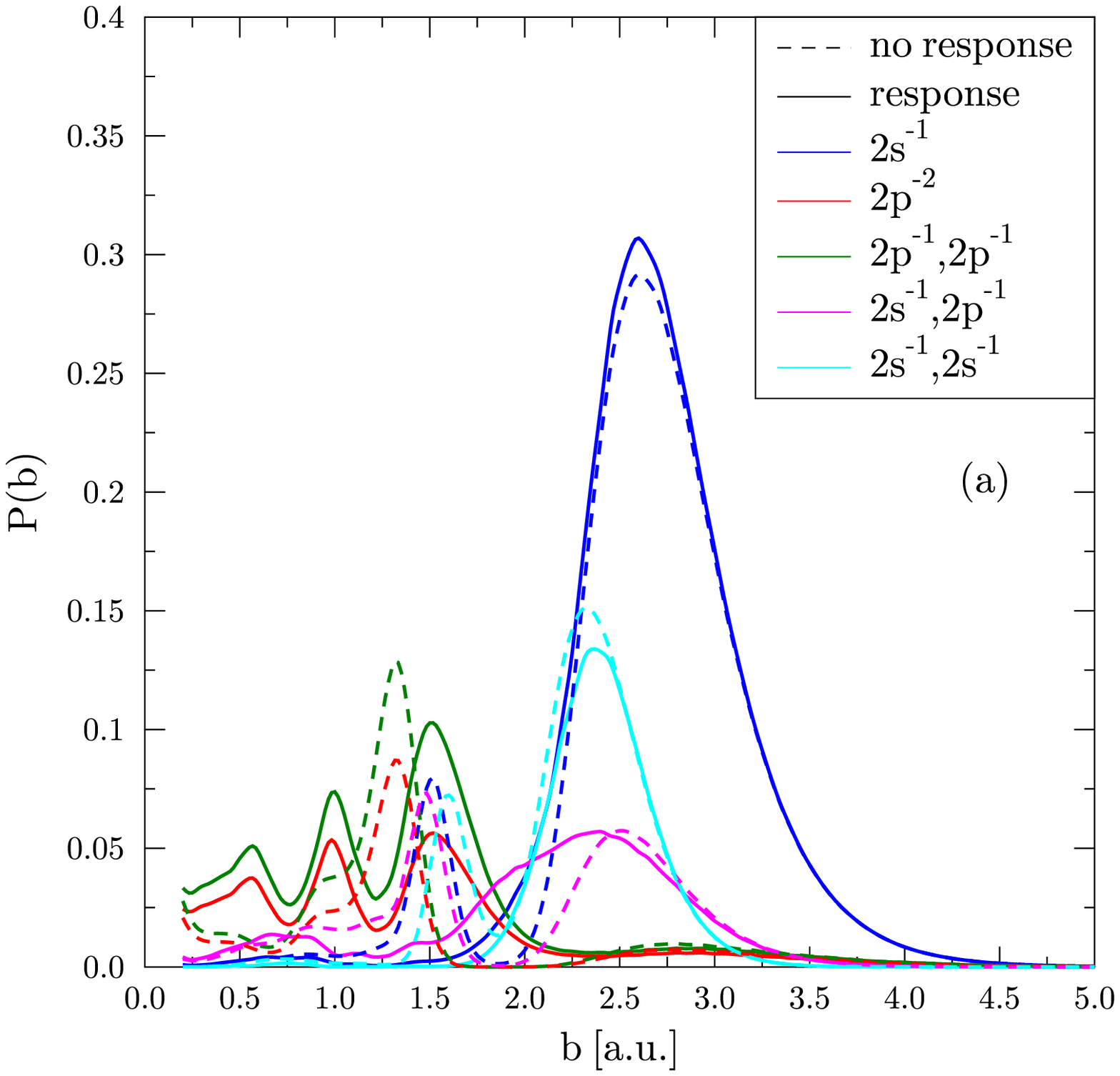}}&
\resizebox{0.52\textwidth}{!}{\includegraphics{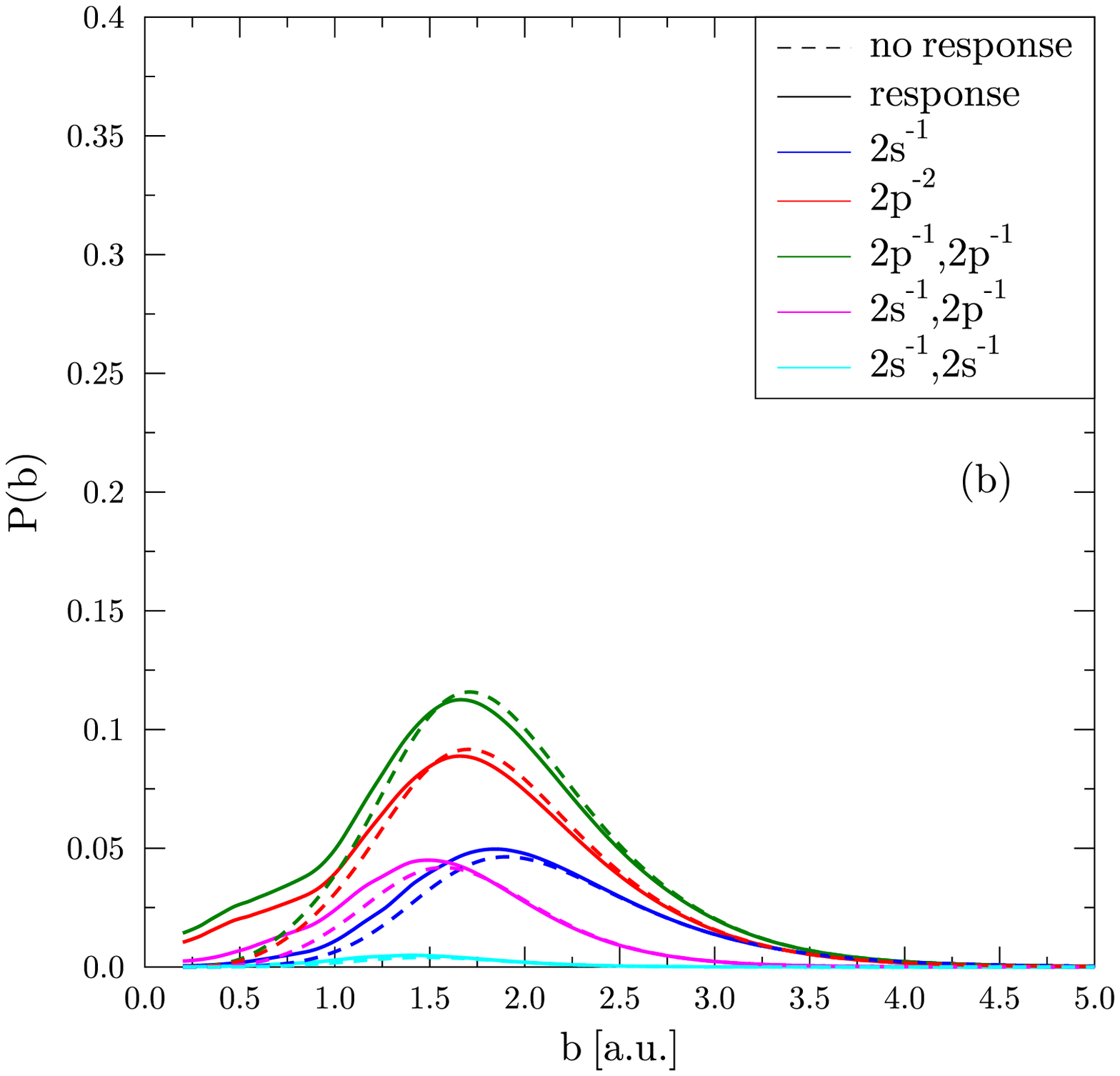}}
\end{array}$
\caption{%
Probabilities for $2s^{-1}$, $2p^{-2}$, ($2p^{-1},2p^{-1}$), ($2s^{-1},2p^{-1}$), and
($2s^{-1},2s^{-1}$) production at
(a) $E_P=10$ keV/amu and (b) $E_P=150$ keV/amu in
He$^{2+}$-Ne$_2$ collisions in orientation I, in which the projectile beam axis is
parallel to the dimer axis,
plotted as functions of the impact parameter. 
}
\label{fig2}
\end{center}
\end{figure}

Figure~\ref{fig2} illustrates for orientation I in which the impact parameters of the atomic and the 
dimer problems are the same, how the single-particle probabilities 
shown in figure~\ref{fig1}
translate into the dimer removal probabilities of interest in this work.
At $E_P=10$ keV/amu the main peak of the $2s$-vacancy production channel associated with
ICD (denoted by $2s^{-1}$ and shown in blue in figure~\ref{fig2}) is almost the same in the no-response and response models despite the seemingly different
single-particle $2s$-removal probabilities. 
However, the $2s^{-1}$ peak occurs at $b=2.6$ a.u., which is where the single-particle
response and no-response probabilities shown
in figure~\ref{fig1}(a) begin to merge. They are strikingly different at smaller $b$ and this is indeed
reflected by larger differences in the $2s^{-1}$ data of figure~\ref{fig2}(a). 
Somewhat counterintuitively, the strong no-response $2s$-removal probability at $b\approx 1.9$ a.u.
(figure~\ref{fig1}(a)) is associated with a close-to-zero $2s^{-1}$ probability of 
the dimer (figure~\ref{fig2}(a)). 
This is a consequence of the multinomial analysis in which the production of exactly one
vacancy in a given subshell is associated with a product of one single-particle
probability for the removal of
an electron and $N-1$ single-particle probabilities for not removing the other 
electrons~(cf.~equation~(\ref{eq:picd})). The latter are very efficient suppression factors.

Clearly, the $2s$-removal process associated with ICD is the strongest channel at
$E=10$ keV/amu. The other two processes involving $2s$ removal (shown in magenta and cyan)
are appreciable as well with their main peaks occurring in the $b=2.3-2.5$ a.u. region.
The processes involving the removal of two or one $2p$ electrons from one or both atoms 
(denoted by $2p^{-2}$ and $(2p^{-1},2p^{-1})$ in the figure) are mostly restricted to 
smaller impact
parameters, which is not surprising given the structure of the single-particle probabilities shown
in figure~\ref{fig1}(a).

At $E=150$ keV/amu the situation is quite different. In this case the $(2p^{-1},2p^{-1})$
and $2p^{-2}$ 
processes dominate,
while the probabilities for the two processes involving the removal of one $2s$ electron 
are smaller and overall of similar
strength, and the process involving the removal of two $2s$ electrons is very weak.
All five probability curves display similar shapes as functions of $b$ which is a direct
consequence of the (mostly) monotonically decreasing behavior of
the single-particle probabilities of figure~\ref{fig1}(b).
Effects due to dynamic response are minor at this energy except for the $2p$ removal
processes at small $b$.

As demonstrated in~\cite{Dyuman20} for triply-charged ion impact, the impact parameter
dependences are markedly different for the other two orientations. Orientation III in which the
dimer is perpendicular to the scattering plane shows mostly small probabilities since the
projectile does not get close enough to either neon atom to cause significant electron removal
except at small $b$.
In orientation II, by contrast, small impact parameter and even head-on ion-atom collisions contribute 
and may result in strong $2p$ electron removal around $b\approx R_e/2 = 2.93$ a.u.
Similarly, for $2s$ vacancy production at $E_P=10$ keV/amu the peak that appears at an impact 
parameter of 2.6 a.u.
in orientation I is effectively moved to the impact parameter $b = 2.6$ a.u. $+ R_e/2 \approx 5.5$ a.u. in
orientation II.

\begin{figure}
\begin{center}$
\begin{array}{cc}
\resizebox{0.52\textwidth}{!}{\includegraphics{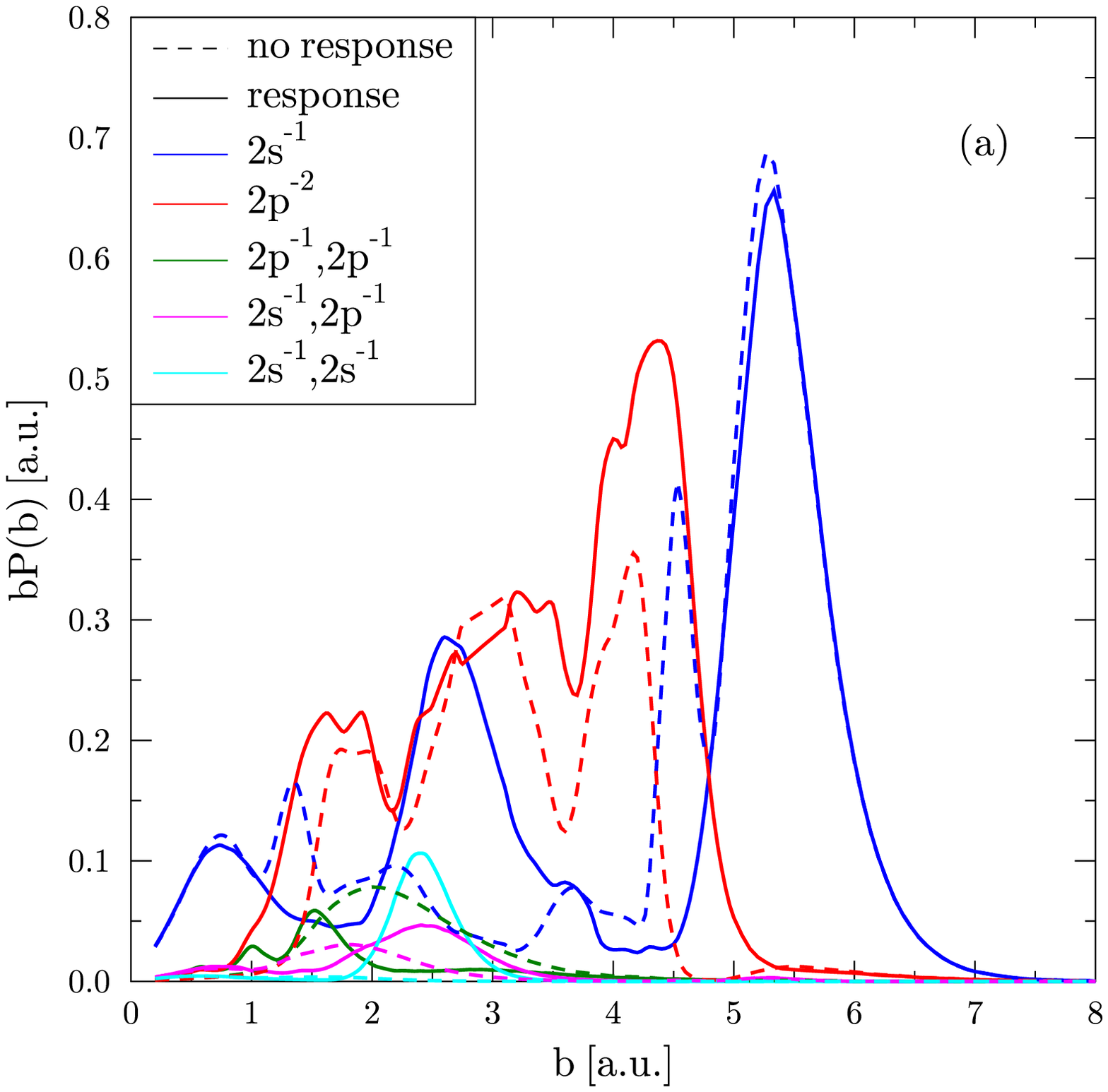}}&
\resizebox{0.52\textwidth}{!}{\includegraphics{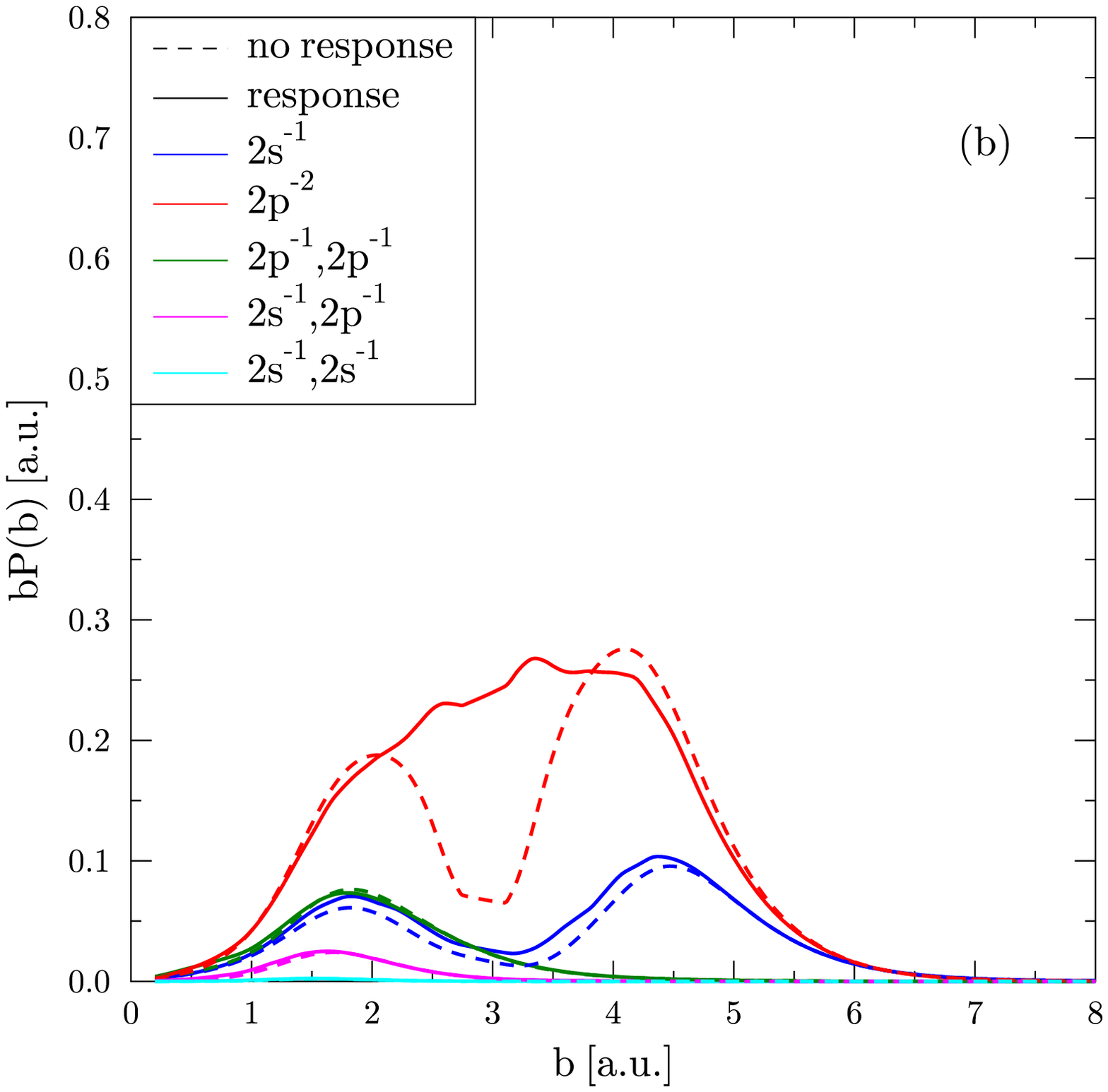}}
\end{array}$
\caption{%
Orientation-averaged impact-parameter-weighted probabilities for $2s^{-1}$, $2p^{-2}$, 
($2p^{-1},2p^{-1}$), ($2s^{-1},2p^{-1}$), and ($2s^{-1},2s^{-1}$) production at
(a) $E_P=10$ keV/amu and (b) $E_P=150$ keV/amu in
He$^{2+}$-Ne$_2$ collisions plotted as functions of the impact parameter. 
}
\label{fig3}
\end{center}
\end{figure}

Figure~\ref{fig3} shows how this all adds up to the orientation-averaged probabilities for the
processes of interest.
In contrast to the previous plots the probabilities shown here are multiplied by the impact
parameter such that the areas under the curves are directly proportional to the 
total cross sections (cf.~equation~(\ref{eq:tcs})).
At $E_P=10$ keV/amu, the $2s^{-1}$ channel is the strongest process in the
no-response model and overall about equally strong as
the $2p^{-2}$ channel when dynamic response is included. 
The other three channels are significantly
weaker. At $E_P=150$ keV/amu the $2p^{-2}$ process prevails in both models. It is stronger
in the dynamic response compared to the no-response model due to contributions around $b=R_e/2$
which originate from close collisions with one of the neon atoms.
The other channels only show small response effects.

\begin{figure}
\begin{center}
\resizebox{0.8\textwidth}{!}{\includegraphics{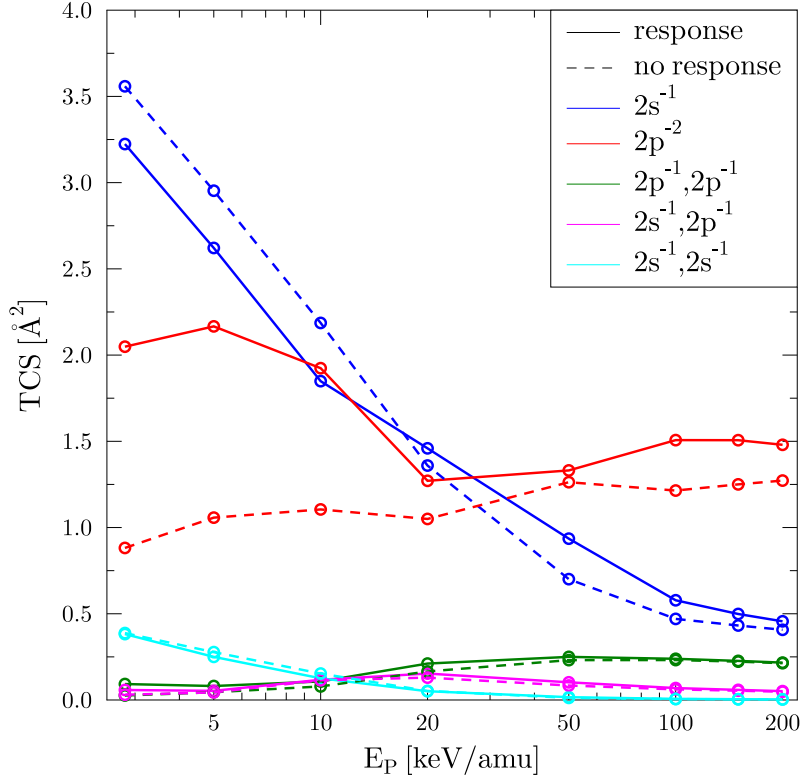}}  
\caption{Orientation-averaged total cross sections for $2s^{-1}$, $2p^{-2}$, ($2p^{-1},2p^{-1}$), 
($2s^{-1},2p^{-1}$), and ($2s^{-1},2s^{-1}$) production in
He$^{2+}$-Ne$_2$ collisions plotted as functions of the impact energy.
}
\label{fig4}
\end{center}
\end{figure}

Figure~\ref{fig4} displays the corresponding total cross sections.
Response effects appear to be strongest for
the $2p^{-2}$ channel, particularly at low energies 
where electron capture dominates. 
To understand the role of response we recall
that (i) in this channel two electrons are removed
from one 
atom, while the other channels involve at most one-electron removal per atom,
and (ii) the response model is designed such that its effects become appreciable
when more than one electron (on average) is removed from an atom.

Regardless of whether the no-response or response models are used, the $2s$-vacancy
production channel associated with ICD is the strongest process in low-velocity collisions.
The lowest energy we consider is $E_P = 2.81$ keV/amu, which is the energy that was
used in the O$^{3+}$-Ne$_2$ experiment reported in~\cite{Iskandar15}.
In that experiment ICD was found to contribute 20\% to the total Ne$^+$ + Ne$^+$ yield,
while CE and RCT accounted for 10\% and 70\%, respectively.
If we follow that work and assume that the $2p^{-2}$ channel leads to RCT in only 50\%
of all instances\footnote{%
Two-electron removal from one of the two neon atoms does not have to be followed by RCT, but
can also result in Ne$^{2+}$ + Ne fragmentation. Since the branching fractions of both
decay modes are unknown, it seems natural to assume a 50:50 breakdown (see also the
analysis presented in~\cite{Iskandar18}).},
while single $2s$-vacancy production is always followed by ICD\footnote{%
ICD in small neon clusters was shown to be at least three orders of magnitude faster
than de-excitation via photon emission~\cite{Santra01} (see also the discussion in
section~4 of the topical review article~\cite{Jahnke15}).},
we obtain for the He$^{2+}$-Ne$_2$ system at $E_P = 2.81$ keV/amu the breakdowns 80:10:10 and
67:11:21 for ICD vs. CE vs. RCT in the no-response and response models, respectively, i.e.
a remarkably strong ICD yield which should be easily identifiable in an experimental
kinetic energy release spectrum.

\begin{figure}
\begin{center}$
\begin{array}{cc}
\resizebox{0.52\textwidth}{!}{\includegraphics{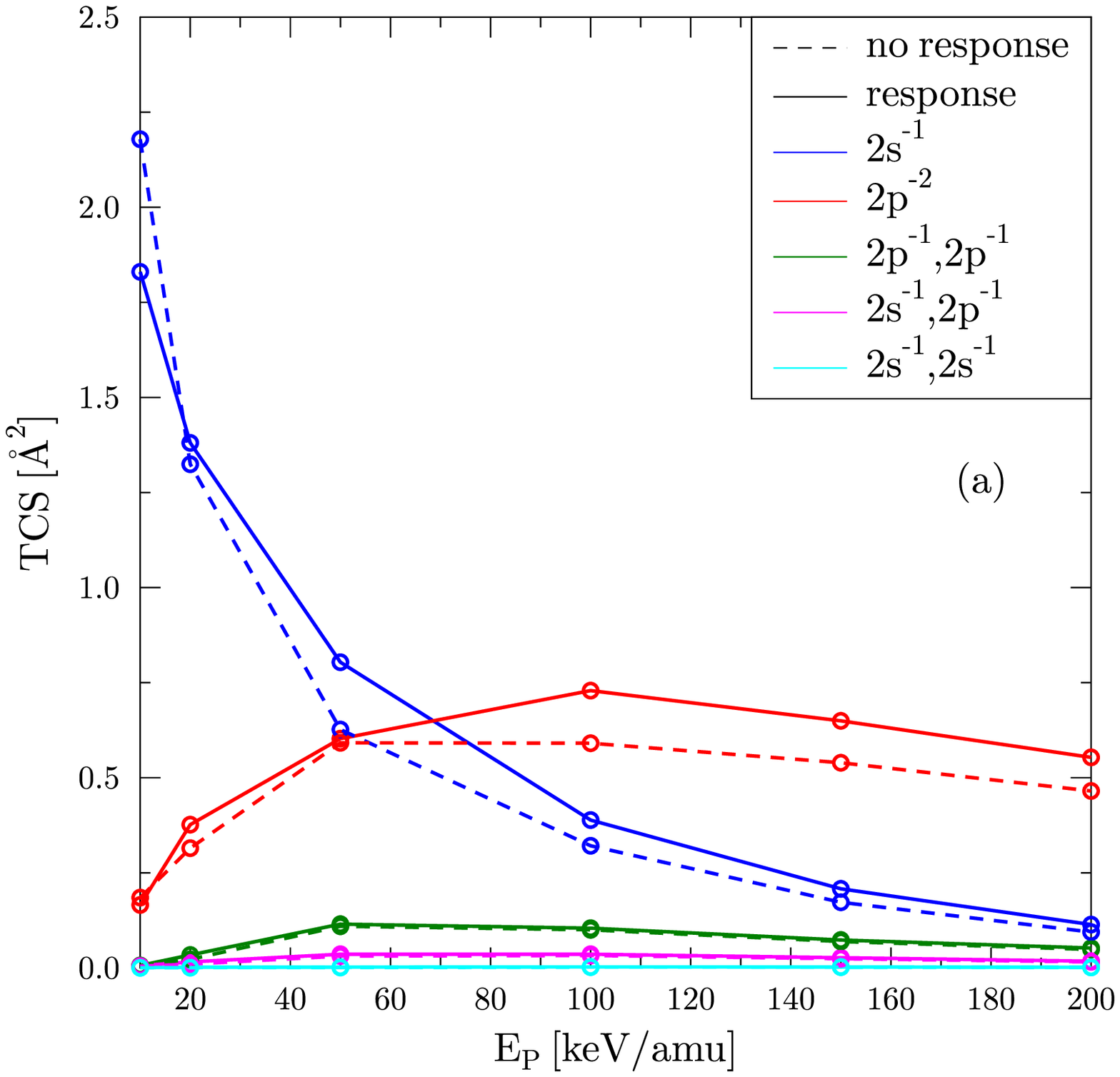}}&
\resizebox{0.52\textwidth}{!}{\includegraphics{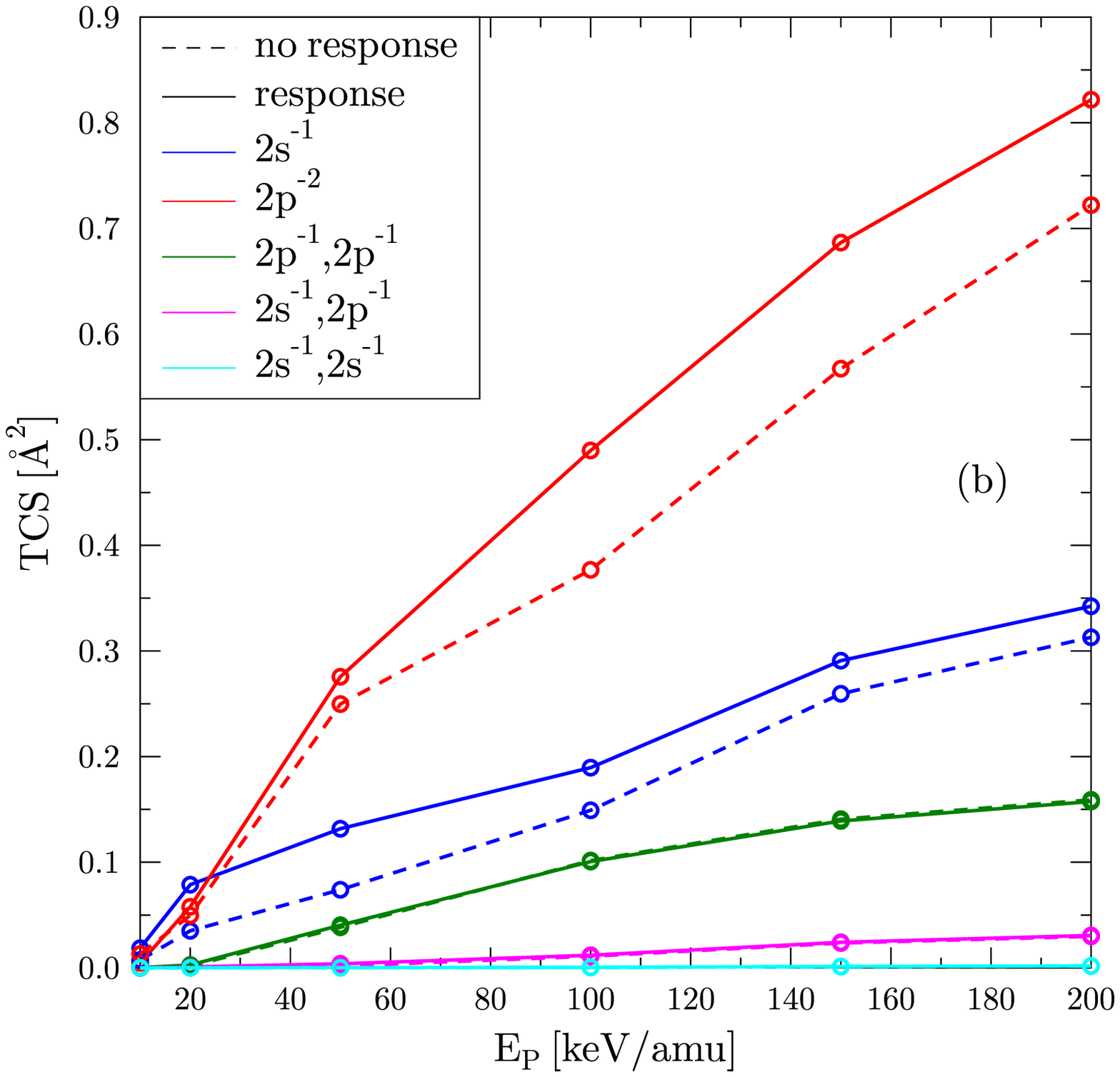}}
\end{array}$
\caption{%
Orientation-averaged total cross sections for $2s^{-1}$, $2p^{-2}$, ($2p^{-1},2p^{-1}$), 
($2s^{-1},2p^{-1}$), and ($2s^{-1},2s^{-1}$) production in
He$^{2+}$-Ne$_2$ collisions plotted as functions of the impact energy for the
(a) transfer ionization channel, (b) double ionization channel (see text for details). 
}
\label{fig5}
\end{center}
\end{figure}

Our results further indicate that ICD remains appreciable (and measurable) toward higher
energies, although RCT takes over as the dominant channel and CE becomes slightly stronger
on a relative scale.
ICD was experimentally identified at $E_P=150$ keV/amu in~\cite{Kim13},
but the relative yields of the three channels were not reported in that work.
Instead, ejected electron spectra were shown for certain channels, and ratios of ICD
vs. direct electron yields extracted from them. To make contact with those measurements we separate
capture and ionization events and show in figure~\ref{fig5}(a) and~\ref{fig5}(b) respectively the partial cross sections that correspond to
finding one electron in the continuum and one electron at the projectile after the collision
(transfer ionization) and to finding two electrons in the continuum (double ionization).
Note that for the $2s^{-1}$ channel we simply look at capture of one $2s$ electron 
and ionization of one
$2s$ electron (cf.~equation~(\ref{eq:picd})) for the transfer ionization and double ionization
channels respectively, 
since our formalism does not describe the ICD continuum electron explicitly.
The plots are terminated at $E_P=10$ keV/amu since the capture vs. ionization separation
becomes inaccurate at lower energies where ionization is very small, and are shown
on linear scales.

The transfer-ionization channel which corresponds to observing singly-charged projectiles
after the collision is the one that was considered in~\cite{Kim13}
at $E_P=150$ keV/amu. We observe in figure~\ref{fig5}(a) that the $2s$-vacancy production
channel has a different shape compared to the others since it does not involve
collisional ionization, but the capture of exactly one $2s$ electron. Accordingly, the
$2s^{-1}$ process is much stronger than the other processes at low impact energies,
but decreases with increasing energy and crosses the $2p^{-2}$ process.
For the double ionization channel shown in figure~\ref{fig5}(b) the shapes of all cross
section curves are similar (monotonically increasing), which reflects the impact-energy
dependence of pure ionization in the energy range considered. The removal of two $2p$
electrons from one atom ($2p^{-2}$) is the strongest process except perhaps at the 
lowest energies where all pure-ionization processes are weak.

\begin{figure}
\begin{center}
\resizebox{0.8\textwidth}{!}{\includegraphics{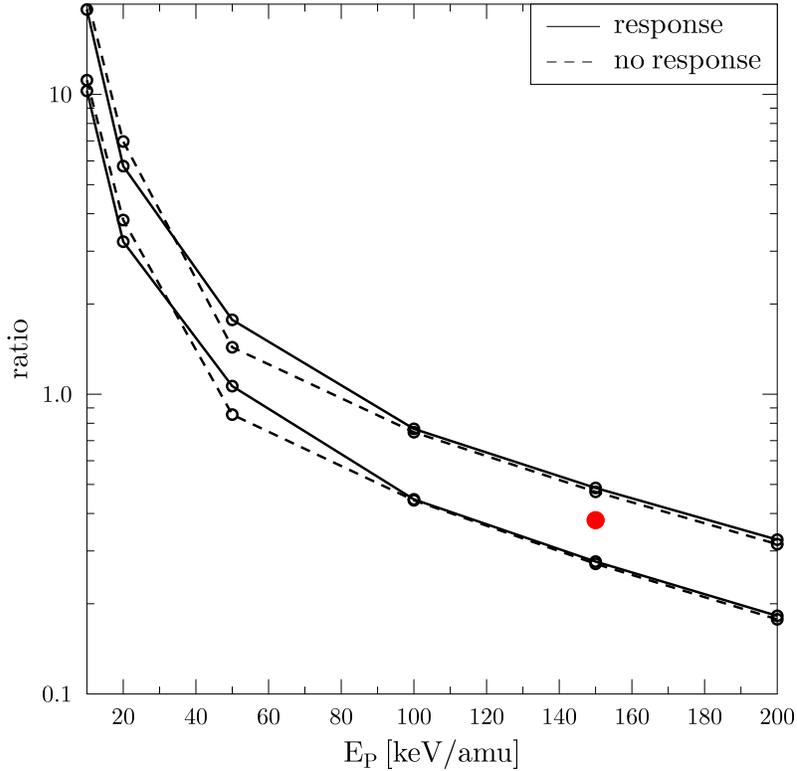}}  
\caption{Cross section ratio of the $2s^{-1}$ process compared to the other four
processes included in figure~\ref{fig5}(a) plotted as a 
function of the impact energy. The lines depict upper and lower bounds obtained by
weighing the $2p^{-2}$ cross section with a factor of 0.5 and 1.0, respectively
(see text for further explanation).
Red bullet: experimental data point from~\cite{Kim13} for the ratio of the
electron yields due to ICD and direct ionization in the transfer ionization
channel.
}
\label{fig6}
\end{center}
\end{figure}

Figure~\ref{fig6} shows the cross section ratio of the $2s^{-1}$ process compared
to the sum of all other processes in the transfer-ionization channel. This should
correspond to the ratio of the total ICD electron yield compared
to the yield of directly ionized electrons for singly-charged helium ions in the
final channel which was measured in~\cite{Kim13}. 
Accordingly, that single experimental data point is included in the figure (using a red
full circle). For both no-response and response models upper and lower bounds of
the ratio are displayed, where the upper bound is obtained from the
above-mentioned assumption that 50\% of the $2p^{-2}$ events result in RCT and the
lower bound from a more extreme 100\% assumption. Response effects are rather small
for this ratio and the experimental data point sits right in the middle between the upper 
and lower bounds (at a value
of 0.38). As can be inferred from figure~\ref{fig5}(a)
the ratio increases to large values between 10 and 20 toward the lowest energies considered.
Once again, this indicates that ICD should be easily detectable 
in a low-energy measurement.

\section{Concluding remarks}
\label{sec:conclusions}
We have studied the one-electron and two-electron processes which are associated with
ICD, CE, and RCT in the Ne$^+$ + Ne$^+$ fragmentation channel in He$^{2+}$-Ne$_2$
collisions. The calculations are at the level of an 
independent-atom-independent-electron model of the neon dimer, and the 
coupled-channel TC-BGM was used to propagate the initially populated orbitals in
the field of the projectile potential. 
In contrast to a previous work for triply-charged projectile ions~\cite{Dyuman20},
both dynamic response and frozen-potential (no-response) calculations were 
carried out, and appreciable differences of both were found, in particular at
low projectile energy. This raises the question if the previous (frozen potential)
results for O$^{3+}$ ion impact, which showed good agreement with available
measurements, would be altered by response effects as well. A quantitative answer
is not straightforward since the presence of electrons on the projectile complicates the
situation for O$^{3+}$ impact. However, two sets of test calculations using slightly
different assumptions for the response potential both indicate that the effects of
response are minor for the relative yields reported in~\cite{Dyuman20}.
Similarly, the relative yields for the ($2s^{-1},2p^{-1}$) and ($2s^{-1},2s^{-1}$)
processes which were not considered in~\cite{Dyuman20} turn out to be very
small, i.e. neither item  changes the conclusions of that work.

Coming back to the He$^{2+}$-Ne$_2$ system studied in the present work, our main
results and conclusions can be summarized as follows: 

(i) Given the simplicity of the response model and the fact that our theoretical description
has other limitations as well~(see below), we are not in a position to conclude that
the response results are necessarily more accurate than the no-response results. Rather, the 
discrepancies between both sets of calculations should perhaps be interpreted more
conservatively as an estimate of the theoretical model uncertainties~\cite{Chung2016}.
Naturally, additional experimental data would help with a more definitive assessment of 
the response vs. no-response results reported here.

(ii) New experimental data may also help assess other limitations of our theoretical 
description~\cite{Dyuman20}, particularly those associated with the 
independent-atom-independent-electron model. 
For example, within this model the dynamical description
of the ion-dimer collision does not take into account
that the projectile charge state changes
when an electron is captured from one of the two neon dimer atoms and that this
change may affect the interaction of the projectile with the other atom. 
Another
process which is ignored is the ion-impact-induced ionization of one of the atoms and 
the subsequent ionization of the other one by a collision with the just released
electron. An estimate using a model described in~\cite{najjari2021} suggests that the 
latter process results in a double ionization cross section
of at most similar magnitude as the cross section for ($2s^{-1},2p^{-1}$) production shown
in figure~\ref{fig5}(b), and it obviously cannot affect the outcome in the 
transfer ionization channel shown in figure~\ref{fig5}(a).
It is more difficult to estimate the former limitation, i.e. effects of a dynamically 
changing projectile charge state 
on the transfer ionization results. Based on experience gathered from ion-atom
collision studies one may argue that single-capture processes might be less affected
than double-capture events~\cite{Schenk15}, but ultimately this can only be decided by
future calculations which take these effects into account.

(iii) With the caveat of the aforementioned model limitations
the main finding of the present work is
that single $2s$-vacancy production is the strongest process at low impact energy,
and, consequently, ICD the dominant
Ne$^+$ + Ne$^+$ fragmentation channel in this region.
This is a remarkable prediction in that it suggests that a large free electron yield
will be observed in the ion-dimer system at collision energies where free-electron
production is a very weak process in the ion-atom problem.

(iv) Key to the dominant role of ICD is strong Ne($2s$) capture with little
simultaneous Ne($2p$) removal. Certainly, the He$^{2+}$-Ne
system will not be unique in this regard, but it appears likely that
this feature is more an exception than the rule for an ion-atom
collision problem. Low-energy collisions are better candidates for giving rise
to strong ICD than collisions at higher energies, since the latter are
dominated by ionization processes which typically increase in strength with
decreasing binding energy, i.e. strong inner-valence electron removal
will be accompanied by even stronger outer-shell electron removal in this regime,
effectively suppressing ICD. 

Given the continuing interest in ICD it is hoped that this work will stimulate
further experimental studies of this process in ion-impact collision problems.
The comparison of accurate measurements and calculations can also shed 
light on the question of which fraction of two-electron removal from one
atom (i.e. the $2p^{-2}$ process) gives rise to RCT. 
Our present and previous results for He$^{2+}$ and O$^{3+}$
collisions suggest that this fraction may be higher than 50\%,
and some of the over-the-barrier model calculations reported in~\cite{Iskandar18} seem
to suggest the same, but a more
definitive answer is outstanding.

\begin{acknowledgments}
Financial support from the Natural Sciences and Engineering Research Council of Canada (NSERC) 
(RGPIN-2019-06305) is gratefully acknowledged. 
\end{acknowledgments}

%

\bibliography{icd}

\end{document}